# Architectural Approaches to Overcome Challenges in the Development of Data-Intensive Systems

**Aleksandar Dimov, Simeon Emanuilov, Boyan Bontchev, Yavor Dankov, and Tasos Papapostolu**

Faculty of Mathematics and Informatics, Sofia University "St Kl. Ohridski", Sofia, Bulgaria

**ABSTRACT**

Orientation of modern software systems towards data-intensive processing raises new difficulties in software engineering on how to build and maintain such systems. Some of the important challenges concern the design of software architecture of such systems. In this article, we survey the fundamental challenges when designing data-intensive computing systems and present some of the most popular software architectural styles together with their potential to tackle these challenges.

# INTRODUCTION

Traditionally, data-intensive systems have emerged from the notion of data-intensive computing, which was inspired by the requirements of traditional science disciplines while moving to the so-called eScience (Hey et al., 2009) – a research perspective that involves extensive usage of Information and Communication Technologies in all research initiatives.

Nowadays, most computing systems collect and process data from various sources, ranges, and applications. As a result, there is a shift in the area of software engineering from software-intensive systems towards data-intensive systems. More challenges reside in size (e.g., amount of data), its complexity, heterogeneity, and velocity. This requires activities that differ from solving traditional software development problems. For example, in data-intensive systems, the machine's computing power is not the main limiting factor but the I/O and additional quality characteristics like reliability, scalability, maintainability, etc. (Kleppmann, 2017).

To manage the level of complexity in software systems and deal with various quality attributes, one should put effort into the design of software architecture. It is widely recognized as an essential factor for the successful development and maintenance of software systems during their entire life-cycle. A common definition of architecture is "the structure or structures of the system, which comprise software components, the externally visible properties of those components and the relationships among them" (Bass et al., 2013). One of the most promising applications of software architecture is the definition of patterns (also called architectural styles) that help to reuse certain solutions to challenges in systems development.

This paper analyzes the applicability of some of the popular architectural styles for addressing the main architectural challenges when designing data-intensive systems.

The structure of the paper is as follows: Section 2 presents the architectural challenges for data-intensive systems; Section 3 discusses some of the most popular architectural styles and makes an analysis of their application with respect to the challenges and finally, Section 4 concludes the paper and states some directions for future research in the area.

## ARCHITECTURAL CHALLENGES FOR BIG DATA

Some research works are available that try to identify architectural challenges in development of data-intensive systems. For example, according to (Mattmann et al., 2011), there exist seven key architectural challenges concerning development and implementation of data-intensive systems:

- Data volumes – huge-scale data volumes, that may go up to petabytes (PB) and exabytes (XB), pose massive problems with transfer and processing compared with average data sizes (e.g., gigabytes – GB). It should be noted that software architecture should also tackle these hardware limitations that may lead to problems caused by data volume.
- Data dissemination – delivery of data to potentially distributed tenants may need to take specific actions when trying to satisfy some specific system qualities like performance, reliability, and security.
- Data curation – architecture should ensure that all data within the data-intensive system share common properties.
- Software reuse and use of open-source software.
- Search of data – the system should provide mechanisms to find and present the location of data to tenants. Note that search is different from data dissemination and retrieval.
- Data processing and analysis – pre- and post- processing/analysis of data in order to be useful for end-users pose new challenges with respect to computational and hardware resources.
- Information modeling – how metadata will be organized to facilitate interoperability and connectivity.

All the above-listed challenges are interconnected with each other. Therefore, the design of the architecture of data-intensive systems and their development requires an overall strategy that addresses all difficulties and challenges.

Furthermore, according to (Zhao et al., 2014), challenges related to the massive amounts of data (Big Data) may also regard the system's scalability, programmability, and performance. As an example, the challenges for data-intensive systems regarding the traditional scientific workflows, involve big data problems referring to "data scale and computation complexity, resource provisioning, collaboration in heterogeneous environments" (Zhao et al., 2014). Concerning the cloud-storage for data-intensive systems, the classification of the main challenges includes (Tudoran et al., 2014):

- Support for massive unstructured data
- Fine-grain access to data subsets
- Support for operations on multiple files
- High throughput under heavy access concurrency
- Support for highly parallel data workflows
- Provision of monitoring and logging services

The first two of these challenges deals with already discussed ones about data volume and data curation, while the next three relate to concurrency and synchronization aspects.

Another kind of challenge also emerges due to data volume – the risk of processing data in a biased way. For example, machine-learning algorithms are a common technique widely used today in numerous fields. But a "blind" application of such algorithms can lead to biased results (Bolukbasi et al., 2016).

Since it is trained on data, such a process that does not consider the specificities of a dataset could amplify the bias existing in the given data. There have been different cases where such algorithms have been found to have biased results and discriminate based on gender, race, etc. The problem is complex and challenging to solve, and only with continuous assessments and testing can be tamed. A positive development that can be identified is that the issue has gained popularity and research interest to eliminate such biased results. A recent example is Twitter (Chowdhury, 2021), which offered a prize for identifying biased results in their image cropping algorithm, and a variety of biased results was pinpointed, such as the favoring of young, thin females or Latin text over Arabic script.

In conclusion of this section, we may distinguish the following main architectural challenges toward database-intensive systems:

- How to deal with big volumes of data, not only about processing, but also in terms of storage, bandwidth, and dissemination.
- How to provide search of data according to end-user requirements (like precision and performance of search).
- How to manage concurrency and provide synchronization and communication between processes and threads of the system.
- How to deal with data curation.
- How to deal with information modeling.

Finally, the main challenges that software architecture should deal with, according to different surveys, may be summarized as follows: (1) Data volume; (2) Data search; (3) Concurrency; (4) Data curation, and (5) Information modeling.

## ARCHITECTURAL STYLES IN DATA INTENSIVE SYSTEMS

A successful generalization of the notion of software architecture is the definition and usage of architectural styles (Garlan, 1994; Mehta, 2003; Sharma

Table 1. Mapping of architectural styles to data-intensive systems challenges.

|  | Data volume | Data search | Concurrency | Data curation | Information modeling |
|---|---|---|---|---|---|
| **Shared data** | E | E | E | N | E |
| **Sharding** | E | E | N | N | N |
| **Pipe and filter** | E | N | E | N | N |
| **Priority queue** | N | N | Y | E | N |
| **Publish-Subscribe** | Y | Y | Y | E | E |
| **Wrapper** | E | E | E | E | E |
| **Caching** | Y | Y | E | N | N |

Legend: Y – Yes; N – No; E – to some extent (needs additional implementation and research).

et al., 2015). They represent successful architectural configurations that recur in different software development projects. Architectural styles define types of software components, the types of connectors between them, and the bindings between them. Finally, styles may be used as standard means to fulfill quality requirements and this way – to resolve typical software design challenges.

In this section, we are going to present some common architectural styles, as they are discussed in various research works in the area of software architecture (Garlan, 1994, Mehta, 2003, Sharma et al. 2015). Along with a brief description of the styles, they will be analyzed with respect to their possibility to tackle the challenges towards data-intensive systems, as described in the previous section (see Table 1 for a summary).

**Shared data** – the essence of this style, as its name says, is the so-called shared data connector that acts as a medium for communication between other components. This is one of the traditional styles used for many years, however, it has new insights concerning data-intensive systems, because when the volume of data increases, it becomes difficult for maintenance. This is further toughened if the shared data is distributed, which is the case of most modern systems. Therefore, the basic implementation of this style is not suitable for application to tackle the challenges for data-intensive software systems, except for reuse, although it is the foundation to deal with data volume, concurrency, and information modeling.

**Sharding** – according to some sharding strategies, this architectural style tries to organize data logically into smaller chunks. The final goal is to optimize access to potentially distributed big data volumes. The sharding style is usually applicable in combination with shared data to deal with large volumes of data and search and dissemination.

**Pipes-and-Filters** – this style has its roots in the dawn of information systems. It defines independent entities called filters (or components) that represent the computational part of the system, and pipes, which serve as connectors between the filters. A single filter can consume or produce data to one or more ports. They can also run concurrently and are not considered directly dependent on each other. A pipe has a single source for its input and a single target for its output. It preserves the sequence of data items and does not alter the data passing through. The most interesting part of this style is the implementation of the pipes. They may represent batch processing or on the other extreme – stream processing units. Most of the modern data-driven architectures and frameworks represent these variations of the pipe-and-filter style. For example, Lambda-architectures (Kraetz, 2021) employ both variations to provide reliable and timely delivery of substantial amounts of data. Other alternatives like Kappa and Delta (Armbrust et al., 2020) architectures also should be considered representatives of this style.

**Priority Queue** – this is another one of the traditional architectural styles, which purpose is to prioritize requests between components in the architecture. This way some of the requests will be processed more quickly (according to predefined or dynamic priorities). Priority queues may serve different purposes mainly in terms of synchronization.

**Publish-Subscribe** – The essence of this style is a common framework (also called a publisher) that acts as a connector between components, which interact via the publisher and have minimal or no knowledge about each other. Variations and earlier implementations of this style are also known as message queuing, event broker, implicit invocation, message passing, or shared bus. This style finds application in modern data-intensive systems (like Kafka) and has been proven to be successful in many aspects. However, it seems to be not suitable in terms of software reuse.

**Wrapper** – is a very general software architectural approach, which aims to "wrap" a component in a way to hide its interfaces from the rest of the world completely. Such an approach may have various goals: better reusability, security, reliability, and so on. Description of particular styles that implement the wrapper approach is beyond the scope of this paper, however, the most popular examples are proxy, broker, adaptor, mediator, façade, ambassador, etc. Wrapper techniques may find application to tackle all challenges of data-intensive systems.

**Caching** – caches are components that may also be classified under the wrapper style, they act as a mediator between the data consumer and the data source, with the latter having larger times to retrieve the data. The goal is to speed up data retrieval as caches provide faster access. Cache is appropriate to deal with in some circumstances with data volumes and search. It can also encourage concurrency if the cache invalidation is properly designed, which needs additional research, however.

## CONCLUSION

The architectural design of data-intensive systems brings many challenges in terms of various systems characteristics and requirements. This paper outlines some essential architectural patterns helping to overcome existing problems in achieving system architectural design that will meet these architectural qualities. The presented patterns are further scrutinized concerning their potential to tackle the previously identified challenges.

Our further work will augment the mapping of architectural styles and patterns to the identified challenges in designing data-intensive systems and applications. Applying the patterns appropriately to such systems, together with the most recent advances and paradigms in software engineering (e.g., the micro-services paradigm), will ensure distributed architectures with high cohesion and very low coupling.

**ACKNOWLEDGMENT**

The research presented in this paper was partially supported by the Sofia University "St. Kliment Ohridski" Research Science Fund Project No. 80-10-74/25.03.2021 – "Data intensive software architectures".